\documentclass[preprint,12pt,fleqn]{elsarticle}

\usepackage{amssymb}
\usepackage{amsfonts}
\usepackage{amsthm}
\usepackage[english]{babel}
\usepackage{euscript}
\usepackage{bbm}
\usepackage{xfrac}
\usepackage{pb-diagram}

\usepackage{mathrsfs}
\usepackage{calrsfs}
\usepackage{amsbsy}

\newcommand{\J}{\mathrm{J}}
\newcommand{\E}{\mathrm{E}}

\usepackage{amssymb}
\usepackage{amsmath}
\usepackage{amsfonts}
\usepackage[english]{babel}
\usepackage{bbm}
\usepackage{xfrac}
\usepackage{color}

\usepackage{mathrsfs}

\usepackage[all]{xy}
\SelectTips{cm}{}

\let\mathcal\mathscr

\newtheorem{theorem}{Theorem}[section]

\newtheorem{proposition}{Proposition}[section]

\let \kappa=\varkappa
\let \phi=\varphi

\newcommand{\abs}[1]{\vert#1\vert}

\begin{document}

\begin{frontmatter}

\title{Lax representations for the 
\\
magnetohydrodynamics equations}

\author{Oleg I. Morozov}   
\ead{oimorozov@gmail.com}
\address{Trapeznikov Institute of Control   Sciences, 65 Profsoyuznaya street, 
\\
Moscow 117997, Russia}

\begin{abstract}
We find two Lax representations for the reduced magnetohydrodynamics equations ({\sc rmhd}) and  derive a 
B{\"a}cklund   transformation between the tangent and the cotangent coverings of these equations. Then, we study the action of the B{\"a}cklund  transformation on the second-order cosymmetries and the action of its inverse on the Lie symmetries of  {\sc rmhd}. 
The  action of the  inverse transformation produces another Lax representation  for {\sc rmhd}.
 The reduction of {\sc rmhd} by the symmetry of shifts along the $z$-axis coincides with the equations of two-dimensional ideal magnetohydrodynamics ({\sc imhd}). Applied to the  Lax representations and the B{\"a}cklund   transformation of {\sc rmhd}, the reduction provides analogous constructions for {\sc imhd}. The action of the inverse B{\"a}cklund   transformation on the Lie symmetries
of {\sc imhd} is expressed in terms of a new four-parameter  Lax representation of these equations. 
  
\end{abstract}


\end{frontmatter}



\section{Introduction}

In this paper, we consider    
 the reduced magnetohydrodynamics equations ({\sc rmhd}) 
\begin{equation}
\left\{
\begin{array}{lc}
\Delta u_t - \Delta v_z - \J(u, \Delta u) + \J(v, \Delta v)=0,& 
\\
v_t-u_z-\J(u, v)=0,&
\end{array}
\right.
\label{RMHD}
\end{equation}
where 
$\Delta u = u_{xx}+u_{yy}$
and 
\begin{equation}
\J(u, v) = u_x\,v_y-u_y\,v_x.
\label{Jacobi_bracket}
\end{equation}  
System \eqref{RMHD} was derived by approximating the three dimensional ideal incompressible magnetohydrodynamics equations with the goal to describe the major features of nonlinear plasma dynamics in tokamaks, \cite{KadomtsevPogutse1974,Strauss1976,Strauss1977}.  The Hamiltonian formulation of {\sc rmhd} was found in  \cite{MorrisonHazeltine1984}, see also   \cite{MarsdenMorrison1984}. The numerical studies revealed impressive ability of the model to simulate tokamak behavior, \cite{CarrerasWadellHicks1979,RosenbluthMonticelloStraussWhite1976}. The extensive research of the Lie symmetry algebra, conservation laws, and invariant solutions for system \eqref{RMHD} has been carried out  in \cite{Samokhin1985,GusyatnikovaSamokhinTitovVinogradovYumaguzhin1989}, see also \cite[Ch. 3, \S 5.4]{VK1999}.

In this paper, we find Lax representations for {\sc rmhd}. Lax representations provide a fundamental framework that enables the application of various techniques for studying nonlinear partial differential equations and are thus considered a key feature indicating their integrability (see
\cite{WE,Zakharov82,RogersShadwick1982,NovikovManakovPitaevskiyZakharov1984,%
Konopelchenko1987,AblowitzClarkson1991,MatveevSalle1991,Olver1993,BacklundDarboux2001}
 and references therein). From the perspective of the geometry of differential equations, Lax representations and other nonlocal structures of integrable nonlinear systems are naturally formulated in the language of differential coverings \cite{KrasilshchikVinogradov1984,KrasilshchikVinogradov1989,VK1999}.

When $v=0$,  system \eqref{RMHD} acquires the form of the two dimensional Euler equation in  the vorticity form, 
\cite[\S 10]{LandauLifshits6},
\begin{equation} 
\Delta u_t = \J(u, \Delta u) 
\label{Euler_eq}
\end{equation}     
with $u_z=0$. In \cite{Li2001}, a Lax representation for this equation was found.  This was generalized in \cite{Morozov2024},  where we  used the technique of twisted extensions of Lie--Rinehart algebras  to derive a family of Lax representations for equa\-ti\-on \eqref{Euler_eq}. This family depends on functional parameters and in\-clu\-des a
 non-re\-mo\-va\-ble spectral parameter.
  
In the present paper we observe that equations \eqref{RMHD} admit the Lax representation \eqref{RMHD_first_covering}, which generalizes the Lax representation for equation \eqref{Euler_eq} found in \cite{Morozov2024}. 
System \eqref{RMHD_first_covering} includes  the  non-removable spectral parameter. 
From \eqref{RMHD_first_covering} we derive the Lax representation \eqref{RMHD_second_covering}.
In  \cite{KrasilshchikMorozov2025}, a B{\"a}cklund transformation  between the tangent and cotangent coverings
of the Euler equation \eqref{Euler_eq} was found. In this paper, we  derive an analogous B{\"a}cklund transformation \eqref{Backlund_transformation} for  equations \eqref{RMHD} and study the actions of this transformation and its inverse 
on the second-order cosymmetries  and the Lie symmetries of these equations. The action of the  inverse transformation
on the scaling symmetry produces another Lax representation \eqref{RMHD_third_covering} for {\sc rmhd}.

When $u_z=v_z=0$, equations \eqref{RMHD} reduce to  the two dimensional ideal magnetohydrodynamics equations  
({\sc imhd}) 
\begin{equation}
\left\{
\begin{array}{lc}
\Delta u_t  - \J(u, \Delta u) + \J(v, \Delta v)=0,& 
\\
v_t-\J(u, v)=0&
\end{array}
\right.
\label{2DiMHD}
\end{equation}
 written in terms of  vorticity $u$ and magnetic potential $v$, see \cite{Zeitlin1992,ZeitlinKambe1993}.  Therefore,  we set $s_z=p_z=q_z=\psi_{1,z}=\psi_{2,z} =\varphi_{1,z} = \varphi_{2,z}= 0$ in systems  \eqref{RMHD_first_covering},  \eqref{RMHD_second_covering}, \eqref{tangent_covering},   \eqref{cotangent_covering}, \eqref{Backlund_transformation} and in Proposition \ref{proposition1}  and obtain the Lax representations \eqref{2DiMHD_covering}, \eqref{2DiMHD_second_covering} and the B{\"a}cklund transformation \eqref{Backlund_transformation_for_2DiMHD} between the tangent and the cotangent coverings  for equations \eqref{2DiMHD}. The careful consideration of the action of the inverse transformation on the symmetries of {\sc imhd} provides its four-parameter Lax representation \eqref{2DiMHD_general_covering}.


\section{Preliminaries and notation}

The presentation in this section closely follows 
\cite{KrasilshchikVinogradov1984,KrasilshchikVinogradov1989,VK1999,KrasilshchikVerbovetskyVitolo2017}. All our considerations are local.

Let $\pi \colon \mathbb{R}^n \times \mathbb{R}^m \rightarrow \mathbb{R}^n$,
$\pi \colon (x^1, \dots, x^n, u^1, \dots, u^m) \mapsto (x^1, \dots, x^n)$, be a trivial bundle, and
$J^\infty(\pi)$ be the bundle of its jets of the infinite order. The local coordinates on $J^\infty(\pi)$ are
$(x^i,u^\alpha,u^\alpha_I)$, where $I=(i_1, \dots, i_n)$ are multi-indices with $i_k \ge 0$, and for every local section
$f \colon \mathbb{R}^n \rightarrow \mathbb{R}^n \times \mathbb{R}^m$ of $\pi$ the corresponding infinite jet
$j_\infty(f)$ is a section $j_\infty(f) \colon \mathbb{R}^n \rightarrow J^\infty(\pi)$ such that
$u^\alpha_I(j_\infty(f))
=\displaystyle{\frac{\partial ^{\abs{I}} f^\alpha}{\partial x^I}}
=\displaystyle{\frac{\partial ^{i_1+\dots+i_n} f^\alpha}{(\partial x^1)^{i_1}\dots (\partial x^n)^{i_n}}}$.
We put $u^\alpha = u^\alpha_{(0,\dots,0)}$. Also, we will simplify notation in the following way: e.g., in the
case of $n=4$, $m=2$ we denote $x^1 = t$, $x^2= x$  $x^3= y$,  $x^4= z$
and $u^1_{(i,j,k,l)}=u_{{t \dots t}{x \dots x}{y \dots y}{z \dots z}}$ with $i$  times $t$, $j$  times $x$, $k$ times $y$,
and $l$ times $z$, $u^2=v$, etc.

The  vector fields
\[
D_{x^k} = \frac{\partial}{\partial x^k} + \sum \limits_{\abs{I} \ge 0} \sum \limits_{\alpha = 1}^m
u^\alpha_{I+1_{k}}\,\frac{\partial}{\partial u^\alpha_I},
\qquad k \in \{1,\dots,n\},
\]
$(i_1,\dots, i_k,\dots, i_n)+1_k = (i_1,\dots, i_k+1,\dots, i_n)$,  are called {\it total derivatives}.
They com\-mu\-te everywhere on $J^\infty(\pi)$.

The {\it evolutionary vector field} associated to an arbitrary vector-valued smooth function
$\varphi \colon J^\infty(\pi) \rightarrow \mathbb{R}^m $ is the vector field
\[
\mathbf{E}_{\varphi} = \sum \limits_{\abs{I} \ge 0} \sum \limits_{\alpha = 1}^m
D_I(\varphi^\alpha)\,\frac{\partial}{\partial u^\alpha_I}
\]
with $D_I=D_{(i_1,\dots\,i_n)} =D^{i_1}_{x^1} \circ \dots \circ D^{i_n}_{x^n}$.

A system of {\sc pde}s 
\begin{equation}
F_r(x^i,u^\alpha_I) = 0
\label{pde}
\end{equation}
 of the order $s \ge 1$ with $\abs{I} \le s$, $r \in \{1,\dots, \rho\}$
for some $\rho \ge 1$, defines the submanifold
$\EuScript{E}=\{(x^i,u^\alpha_I)\in J^\infty(\pi)\,\,\vert\,\,D_K(F_r(x^i,u^\alpha_I))=0,\,\,\abs{K}\ge 0\}$
in $J^\infty(\pi)$.

A function $\varphi \colon J^\infty(\pi) \rightarrow \mathbb{R}^m$ is called a {\it (generator of an
infinitesimal) symmetry} of equation $\EuScript{E}$ when $\mathbf{E}_{\varphi}(F) = 0$ on $\EuScript{E}$. The
symmetry $\varphi$ is a solution to the {\it defining system}
\begin{equation}
\ell_{\EuScript{E}}(\varphi) = 0
\label{defining_eqns} 
\end{equation}
of equation $\EuScript{E}$, where $\ell_{\EuScript{E}} = \ell_F \vert_{\EuScript{E}}$ with the matrix differential operator
\[
\ell_F = \left(\sum \limits_{\abs{I} \ge 0}\frac{\partial F_r}{\partial u^\alpha_I}\,D_I\right).
\]
The {\it symmetry algebra} $\mathrm{sym} (\EuScript{E})$ of equation $\EuScript{E}$ is the linear space of
solutions to  (\ref{defining_eqns}) endowed with the structure of a Lie algebra over $\mathbb{R}$ by the
{\it Jacobi bracket} $\{\varphi,\psi\} = \mathbf{E}_{\varphi}(\psi) - \mathbf{E}_{\psi}(\varphi)$.
The {\it algebra of contact symmetries} $\mathrm{sym}_0 (\EuScript{E})$ is the Lie subalgebra of
$\mathrm{sym} (\EuScript{E})$ defined as $\mathrm{sym} (\EuScript{E}) \cap C^\infty(J^1(\pi))$.

A {\it cosymmetry} of equation $\EuScript{E}$ is a solution $\psi$ to the system
$\ell^{*}_{\EuScript{E}}(\psi) = 0$,
where $\ell^{*}_{\EuScript{E}}$ is the adjoint operator to the operator $\ell_{\EuScript{E}}$. Cosymmetries are 
an  effective tool for computing conservation laws of {\sc pde}s.

Let the linear space $\EuScript{W}$ be either $\mathbb{R}^N$ for some $N \ge 1$ or  $\mathbb{R}^\infty$
endowed with  local co\-or\-di\-na\-tes $w^a$, $a \in \{1, \dots , N\}$ or  $a \in  \mathbb{N}$, respectively.
The variables $w^a$ are cal\-led {\it pseudopotentials} \cite{WE}.  Locally, a {\it differential covering} of $\EuScript{E}$ is 
a trivial bundle $\varpi \colon J^\infty(\pi) \times \EuScript{W} \rightarrow J^\infty(\pi)$ equipped with the {\it extended total derivatives}
\[
\widetilde{D}_{x^k} = D_{x^k} + \sum \limits_{ s =0}^\infty
T^s_k(x^i,u^\alpha_I,w^j)\,\frac{\partial }{\partial w^s}
\]
such that $[\widetilde{D}_{x^i}, \widetilde{D}_{x^j}]=0$ for all $i \not = j$ on $ \EuScript{E}$.
Define the par\-ti\-al derivatives of $w^s$ by  $w^s_{x^k} =  \widetilde{D}_{x^k}(w^s)$.  This gives the over-determined system
of {\sc pde}s
\begin{equation}
w^s_{x^k} = T^s_k(x^i,u^\alpha_I,w^j), 
\label{WE_prolongation_eqns}
\end{equation}
which is compatible whenever $(x^i,u^\alpha_I) \in \EuScript{E}$.
System \eqref{WE_prolongation_eqns} is referred to as the {\it covering equations} or the {\it Lax representation} of 
equation $\EuScript{E}$. 

The {\it tangent covering} and the {\it cotangent covering} are defined by appending the equations 
$\ell_{\EuScript{E}}(q) = 0$ and $\ell^{*}_{\EuScript{E}}(p) = 0$, respectively, to equations \eqref{pde}.

Two differential coverings $\varpi_1$ and $\varpi_2$ of equation $\EuScript{E}$ 
with the extended total derivatives $\widetilde{D}_{x^k}^{(1)}$ 
and $\widetilde{D}_{x^k}^{(2)}$ 
are called {\it equivalent} if  there exists a diffeomorphism $\Phi$ such that the diagram  
\[
\begin{diagram}
\node{\EuScript{E}\times \EuScript{W}_1}
     \arrow[2]{e,t}{\Phi}
     \arrow{se,r}{\varpi_1\vert_{\EuScript{E}}}
\node[2]{\EuScript{E}\times \EuScript{W}_2}
      \arrow{sw,r}{\varpi_2\vert_{\EuScript{E}}}                            
\\
\node[2]{\EuScript{E}
}
\end{diagram}
\]
is commutative and $\Phi_{*}(\widetilde{D}_{x^k}^{(1)})=\widetilde{D}_{x^k}^{(2)}$.

Let an equation $\tilde{\EuScript{E}}$ cover equations $\EuScript{E}_1$ and $\EuScript{E}_2$,
\[
\begin{diagram}
\node[2]{\tilde{\EuScript{E}}}
     \arrow{sw,l}{\varpi_1}
       \arrow{se,l}{\varpi_2}
\\
\node{\EuScript{E}_1}                                 
\node[2]{\EuScript{E}_2;}
\end{diagram}
\]
then we say that above diagram is a {\it B{\"a}cklund transformation} between 
$\EuScript{E}_1$ and $\EuScript{E}_2$.


\section{Lax representations for RMHD}

The natural setting for {\sc rmhd}  and its Lax representations can be expressed in terms of the Poisson algebra
$\mathfrak{p} = C^\infty(M)/\mathbb{R}$ of non-constant functions on an open subset $M \subseteq \mathbb{R}^2$, equipped with the local coordinates $(x,y)$ and the Jacobi bracket  \eqref{Jacobi_bracket}. The map 
$\iota \colon \mathfrak{p} \rightarrow \mathfrak{h}$, $\iota \colon f \mapsto - f_y\,\partial_x + f_x\,\partial_y$
establishes an isomorphism of $\mathfrak{p}$ and the Lie algebra 
$\mathfrak{h}= \mathrm{sdiff}(M)  = \{ V \in TM \,\,\vert\,\, L_V\,(dx \wedge dy) = 0\}$
of the volume-preserving vector fields  on $M$. The Lie algebra $\mathfrak{h}$ admits the extension by the outer derivation 
$V_0 \in H^1(\mathfrak{h},\mathfrak{h})$, where $V_0$ is a vector field on $M$ such that 
$L_{V_0} \,(dx \wedge dy)= dx \wedge dy$. We can take $V_0 =x\,\partial_x+y\,\partial_y$, then we have  
$[V_0, \iota (f)] = \iota (\E(f))$, where the differential operator 
$\E \in H^1(\mathfrak{p}, \mathfrak{p})$ has the form $\E(f)=x\,f_x+y\,f_y -2\,f$.

We  employ the Lie algebra $\tilde{\mathfrak{p}} = C^\infty(\mathbb{R}^2,\mathfrak{p})$
of smooth functions of $(t,z) \in \mathbb{R}^2$ taking  values in  the Lie algebra $\mathfrak{p}$, endowed with the pointwise 
bracket  $\J$, and consider the functions $u$, $v$  and the pseudopotentials $s$, $p$, $q$  below as the elements of 
$\tilde{\mathfrak{p}}$. 

We have the following observation.

\begin{theorem}
\label{theorem1}
The system
\begin{equation}
\left\{
\begin{array}{lcl}
s_t &=& \J(u, s)+\Delta v + \lambda\,\E(u),
\\
s_z &=& \J(v, s)+\Delta u+\lambda\,\E(v),
\end{array}
\right.
\label{RMHD_first_covering}
\end{equation}
 provides a Lax representation for  {\sc rmhd}.
The parameter $\lambda$ 
 is non-removable,  that is, differential coverings defined by system \eqref{RMHD_first_covering} with different constant values of 
$\lambda$ are not equivalent.
\end{theorem}
\noindent
Proof.  We have 
$(s_t)_z-(s_z)_t = -F_1 - \J(F_2, s) - \lambda\,\mathrm{E}(F_2)$,
 where $F_1$ and $F_2$ are the left-hand sides of  equations \eqref{RMHD}, respectively. Hence the compatibility condition 
$(s_t)_z-(s_z)_t =0$ follows from \eqref{RMHD}.

The symmetry $W=x\,\partial_x+y\,\partial_y +2\,u\,\partial_u+2\,v\,\partial_v$ of equations \eqref{RMHD} does not admit a lift to a symmetry of equations \eqref{RMHD_first_covering} with $\lambda \neq 0$. The action of the prolongation of the diffeomorphism 
$\mathrm{exp}(\varepsilon \, W)$ to the bundle $J^3(\varpi)$ of the third-order 
jets of the sections of the bundle $\varpi \colon \mathbb{R}^7 \rightarrow \mathbb{R}^4$, 
$\varpi \colon (t,x,y,z,u,v,s) \mapsto (t,x,y,z)$,   maps   equations 
\eqref{RMHD_first_covering} with $\lambda =1$ to the same equations with $\lambda =\mathrm{e}^\varepsilon$.  
Therefore the second assertion follows from \cite[\S\S~3.2, 3.6]{KrasilshchikVinogradov1989},
\cite{Krasilshchik2000,IgoninKrasilshchik2000,IgoninKerstenKrasilshchik2002,Marvan2002}.
\hfill $\Box$

\vskip 7 pt

Let
$s_1$, $s_2$, and $s_3$ be solutions to  \eqref{RMHD_first_covering} with the same values of $u$ and $v$. 
Then, $p=s_1-s_2$ and $q=\J(s_3,s_1-s_2)$ are solutions to the system 
\begin{equation}
\left\{
\begin{array}{lcl}
p_t &=& \J(u, p),
\\
p_z &=& \J(v, p),
\\
q_t &=& \J(u, q) + \J(\Delta v+\lambda\,\E(u), p),
\\
q_z &=& \J(v, q) + \J(\Delta u+\lambda\,\E(u), p).
\end{array}
\right.
\label{RMHD_second_covering}
\end{equation}
The proof of the following theorem  is similar to that of Theorem  \ref{theorem1}. 

\begin{theorem} 
\label{theorem2}
 System \eqref{RMHD_second_covering}
defines a Lax representation for system \eqref{RMHD}.  The parameter $\lambda$ in this system  is non-removable.
\hfill $\Box$
\end{theorem}

\vskip 5 pt
\noindent
{\sc Remark}.
Equation \eqref{RMHD} and its Lax representation \eqref{RMHD_third_covering}  allow a natural formulation in terms of the 
Lie al\-ge\-bra $\mathfrak{g}= \mathbb{R}_1[h] \otimes \tilde{\mathfrak{p}}$, where 
$\mathbb{R}_1[h] = \mathbb{R}[h]/\langle h^2=0\rangle$  is the commutative  associative unital al\-ge\-bra of the  truncated polynomials of degree at most 1 in the formal variable  $h$ with $h_t=h_x=h_y=h_z=0$ and $h^k =0$ for $k>1$. As a vector space, 
$\mathfrak{g} = \tilde{\mathfrak{p}}  \oplus \tilde{\mathfrak{p}}$ is the collection of the pairs of smooth functions from 
$\mathbb{R}^2$ to  $\mathfrak{p}$. This vector space is equipped  with the bracket  
$\J(f_0+h\,f_1,g_0+h\,g_1) =\J(f_0,g_0)+h\,(\J(f_0,g_1)+\J(f_1,g_0))$. 
Then equation \eqref{RMHD} and system \eqref{RMHD_third_covering}  can be written in the forms
\[
(v+h\,\Delta u)_t-(u+h\,\Delta v)_z -\J(u+h\,\Delta v, v+h\,\Delta u)=0
\]
and 
\[
\left\{
\begin{array}{lcl}
(p+h\,q)_t &=& \J(u+h\,(\Delta v+\lambda\,\E(u)), p+h\,q),
\\
(p+h\,q)_z &=& \J(v+h\,(\Delta u+\lambda\,\E(v)), p+h\,q),
\end{array}
\right.
\]
respectively,
with $p+h\,q, u+h\,(\Delta v+\lambda\,\E(u)), v+h\,(\Delta u +\lambda\,\E(v))      \in  \mathfrak{g}$.
\hfill $\diamond$


\section{A B{\"a}cklund transformation between the tangent and the cotangent coverings}

In \cite{KrasilshchikMorozov2025}, a  B{\"a}cklund transformation between the tangent and the cotangent coverings
of Euler's equation \eqref{Euler_eq} is presented. It turns out that an ana\-lo\-go\-us transformation 
can be constructed for {\sc rmhd}.

The tangent covering and the cotangent covering of {\sc rmhd}
are obtained by appending the systems
\begin{equation}
\left\{
\begin{array}{rc}
\Delta \varphi_{1,t}-\Delta \varphi_{2,z} - \J (\varphi_1, \Delta u)-\J (u,\Delta \varphi_1)&
\\
+ \J(\varphi_2, \Delta v)+\J (v,\Delta \varphi_2) &=0,
\\
\varphi_{1,z}-\varphi_{2,t} -\J(\varphi_1,v)-\J(u,\varphi_2)&=0
\end{array}
\right.
\label{tangent_covering}
\end{equation}
and
\begin{equation}
\left\{
\begin{array}{rc}\Delta (\psi_{1,t} -\J(u,\psi_1))-\psi_{2,z}+\J(v, \psi_2)+\J(\Delta u, \psi_1)&=0, 
\\
\Delta (\psi_{1,z} -\J(v,\psi_1))-\psi_{2,t}+\J(u, \psi_2)+\J(\Delta v, \psi_1)&=0,
\end{array}
\right.
\label{cotangent_covering}
\end{equation}
respectively, to equations \eqref{RMHD}.
A direct computation proves the following sta\-te\-ment. 

\begin{proposition}
\label{proposition1}
Let $(\psi_1,\psi_2)$ be a solution to system \eqref{cotangent_covering}. Then 
the pair $(\psi_{1,t}-\J(u,\psi_1), \psi_{1,z}-\J(v,\psi_1))$ is a solution to system\eqref{tangent_covering}.
\hfill $\Box$
\end{proposition}

Substituting for  $\psi_{1,t}=\J(u,\psi_1)+\varphi_1$ and  $\psi_{1,z}=\J(v,\psi_1)+\varphi_2$ into 
\eqref{tangent_covering}
gives the following system
\begin{equation}
\left\{
\begin{array}{lcl}
\psi_{1,t} &=&\J(u,\psi_1)+\varphi_1, 
\\
\psi_{1,z} &=&\J(v,\psi_1)+\varphi_2,
\\
\psi_{2,t} &=& \J(u,\psi_2)+\J(\Delta v, \psi_1)+\Delta \varphi_2,
\\ 
\psi_{2,z} &=& \J(v,\psi_2)+\J(\Delta u, \psi_1)+\Delta \varphi_1.
\end{array}
\right.
\label{Backlund_transformation}
\end{equation}
The compatibility conditions $(\psi_{i,t})_z=(\psi_{i,z})_t$ lead to the following result:
\begin{theorem}
System \eqref{Backlund_transformation} defines a B{\"a}cklund transformation between the 
tangent covering and the cotangent covering of {\sc rmhd}. 
\hfill $\Box$
\end{theorem}

Further, we study the action  $\EuScript{H} \colon (\psi_1,\psi_2) \mapsto (\varphi_1,\varphi_2)$  of  the B{\"a}cklund 
trans\-for\-ma\-ti\-on \eqref{Backlund_transformation} on the collection $\mathrm{cosym}_0(\EuScript{E}_1)$ of  the second-order cosymmetries  and  the action of its inverse  $\EuScript{H}^{-1} \colon (\varphi_1,\varphi_2)  \mapsto (\psi_1,\psi_2)$ on the Lie algebra $\mathrm{sym}_0(\EuScript{E}_1)$ of the contact symmetries of equation \eqref{RMHD}. We note that 
$\EuScript{H}^{-1}$ acts nontrivially on the trivial symmetry $(0,0)$, which follows immediately from
system \eqref{Backlund_transformation} with $(\varphi_1, \varphi_2) =(0, 0)$. In this case, system 
\eqref{Backlund_transformation} reduces to system \eqref{RMHD_second_covering} with $\lambda=0$.    
Therefore, we  consider the action of $\EuScript{H}^{-1}$ modulo a so\-lu\-ti\-on of  \eqref{RMHD_second_covering} with 
$\lambda=0$.

The algebra $\mathrm{sym}_0(\EuScript{E}_1)$  is generated by the functions
\[
\begin{array}{lcl}
\Phi_0 &=&(-x\,u_x-y\,u_y+2\,u, -x\,v_x-y\,v_y+2\,v), 
\\
\Phi_1(A_1) &=& (-A_{1,z}\,u_t-A_{1,t}\,u_z-A_{1,tz}\,u-A_{1,tt}\,v, 
\\&&\qquad
 -A_{1,z}\,v_t-A_{1,t}\,v_z-A_{1,tt}\,u-A_{1,tz}\,v),
\\
\Phi_2(A_2) &=& (2\,A_2\,(y\,u_x-x\,u_y)-A_{2,t}\,(x^2+y^2),
\\&&\qquad
 2\,A_2\,(y\,v_x-x\,v_y)-A_{2,z}\,(x^2+y^2)),
\\
\Phi_3(A_3) &=& (-A_3\,u_y-A_{3,t}\,x,-A_3\,v_y-A_{3,z}\,x),
\\
\Phi_4(A_4) &=& (-A_4\,u_x+A_{4,t}\,y, -A_4\,v_x+A_{4,z}\,y),
\\
\Phi_5(A_5) &=& (A_{5,t}, A_{5,z}),
\end{array}
\] 
while  the linear space  $\mathrm{cosym}_0(\EuScript{E}_1)$ is spanned  by the cosymmetries 
\[
\begin{array}{lcl}
\Psi_1(B_1) &=& (B_{1,z}\,u+B_{1,t}\,v, B_{1,t}\,\Delta u+ B_{1,z}\,\Delta v),
\\
\Psi_2(B_2) &=& (B_{2,z}\,(x^2+y^2), 4 B_{2,t}),
\\
\Psi_3(B_3) &=& (B_3\,x,0),
\\
\Psi_4(B_4) &=& (B_4\,y,0),
\\
\Psi_5(B_5) &=& (B_5, 0),
\end{array}
\] 
where $A_i$ and $B_i$ are smooth functions of $t$ and $z$ subject to the restrictions $A_{j,zz}=A_{j,tt}$, $B_{j,zz}=B_{j,tt}$ when $j \in \{1,2\}$. A discussion of the structure of the Lie algebra $\mathrm{sym}_0(\EuScript{E}_1)$, the set 
$\mathrm{cosym}_0(\EuScript{E}_1)$, and their applications to finding invariant solutions and conservation laws of {\sc rmhd} can be found in \cite{Samokhin1985,GusyatnikovaSamokhinTitovVinogradovYumaguzhin1989} and \cite[Ch. 3, \S 5.4]{VK1999}.

The action $\EuScript{H}\colon \mathrm{cosym}_0(\EuScript{E}_1) \rightarrow \mathrm{sym}_0(\EuScript{E}_1)$ 
is given by the formulas 
\[
\begin{array}{lcl}
\EuScript{H}(\Psi_1(A_1)) &=& -\Phi_1(A_1),
\\
\EuScript{H}(\Psi_2(A_2)) &=& -\Phi_2(A_{2,z}),
\\
\EuScript{H}(\Psi_3(A_3)) &=& -\Phi_3(A_3),
\\
\EuScript{H}(\Psi_4(A_4)) &=& \Phi_4(A_4),
\\
\EuScript{H}(\Psi_5(A_5)) &=& \Phi_5(A_5).
\end{array}
\]
These formulas define the local expressions for the action of the inverse operator $\EuScript{H}^{-1}$ on the symmetries 
$\Phi_i(A_i)$ with $i \in \{1, \dots, 5\}$, while the action on $\Phi_0$ is given by $\EuScript{H}^{-1}(\Phi_0) = (p_0,q_0)$,
where $(p_0,q_0)$ is a solution to system 
\begin{equation}
\left\{
\begin{array}{lcl}
p_t &=& \J(u, p)-\E(u),
\\
p_z &=& \J(v, p)-\E(v),
\\
q_t &=& \J(u, q) + \J(\Delta v, p)-\E(\Delta v),
\\
q_z &=& \J(v, q) + \J(\Delta u, p)-\E(\Delta u).
\end{array}
\right.
\label{RMHD_third_covering}
\end{equation}

\section{The two-dimensional ideal magnetohydrodynamics equation}
Upon restricting  $u_z=v_z=s_z=s_z=p_z=q_z=\psi_{1,z}=\psi_{2,z} =\varphi_{1,z} = \varphi_{2,z}= 0$ in systems  \eqref{RMHD_first_covering} and \eqref{RMHD_second_covering}, we obtain the following result. 
\begin{theorem}  
The systems
\begin{equation}
\left\{
\begin{array}{rcl}
s_t &=& \J(u, s)+\Delta v + \lambda\,\mathrm{E}(u),
\\
0 &=& \J(v,s)+\Delta u+\lambda\,\mathrm{E}(v)
\end{array}
\right.
\label{2DiMHD_covering} 
\end{equation}
and 
\begin{equation}
\left\{
\begin{array}{lcl}
p_t &=& \J(u, p),
\\
0 &=& \J(v, p),
\\
q_t &=& \J(u, q) + \J(\Delta v+\lambda\,\E(u), p),
\\
0 &=& \J(v, q) + \J(\Delta u+\lambda\,E(v), p)
\end{array}
\right.
\label{2DiMHD_second_covering}
\end{equation}
provide  Lax representations for system \eqref{2DiMHD}. The parameter $\lambda$ in these systems is non-removable.
\hfill $\Box$
\end{theorem}

Likewise, for the tangent and the cotangent coverings of equation \eqref{2DiMHD},  defined by the equations 
\begin{equation}
\left\{
\begin{array}{rc}
\Delta \varphi_{1,t} - \J (\varphi_1, \Delta u)-\J (u,\Delta \varphi_1)
+ \J(\varphi_2, \Delta v)+\J (v,\Delta \varphi_2) &=0,
\\
\varphi_{2,t} +\J(\varphi_1,v)+\J(u,\varphi_2)&=0
\end{array}
\right.
\label{tangent_covering_of_2DiMHD}
\end{equation}
and
\begin{equation}
\left\{
\begin{array}{rc}\Delta (\psi_{1,t} -\J(u,\psi_1))+\J(v, \psi_2)+\J(\Delta u, \psi_1)&=0, 
\\
-\Delta (\J(v,\psi_1))-\psi_{2,t}+\J(u, \psi_2)+\J(\Delta v, \psi_1)&=0,
\end{array}
\right.
\label{cotangent_covering_of_2DiMHD}
\end{equation}
respectively, the following theorem holds.

\begin{theorem}
Let $(\psi_1,\psi_2)$ be a solution to system \eqref{cotangent_covering_of_2DiMHD}. Then 
the pair $(\psi_{1,t}-\J(u,\psi_1), \psi_{1,z}-\J(v,\psi_1))$ is a solution to system \eqref{tangent_covering_of_2DiMHD}.
The system
\begin{equation}
\left\{
\begin{array}{lcl}
\psi_{1,t} &=&\J(u,\psi_1)+\varphi_1, 
\\
0 &=&\J(v,\psi_1)+\varphi_2,
\\
\psi_{2,t} &=& \J(u,\psi_2)+\J(\Delta v, \psi_1)+\Delta \varphi_2,
\\ 
0 &=& \J(v,\psi_2)+\J(\Delta u, \psi_1)+\Delta \varphi_1
\end{array}
\right.
\label{Backlund_transformation_for_2DiMHD}
\end{equation} defines a B{\"a}cklund \,transformation 
between the tangent covering and the cotangent covering of equation \eqref{2DiMHD}.
\hfill $\Box$
\end{theorem}
The algebra $\mathrm{sym}_0(\EuScript{E}_2)$ of equation \eqref{2DiMHD} is generated by the functions
\[
\begin{array}{lcl}
\varphi_0 &=&(-x\,u_x-y\,u_y+2\,u, -x\,v_x-y\,v_y+2\,v), 
\\
\varphi_1 &=& (-t\,u_t-u, -t\,v_t-v),
\\
\varphi_2 &=& (t\,(x\,u_y-y\,u_x)+\frac{1}{2}\,(x^2+y^2),
t\,(y\,v_x-x\,v_y)),
\\
\varphi_3 &=& (0, 1),
\\
\varphi_4 &=& (-u_t, -v_t),
\\
\varphi_5&=& (x\,u_y-y\,u_x, x\,v_y-y\,v_x),
\\
\Phi_1(A_1) &=& (-A_1\,u_x+A_1^{\prime}\,y,-A_1\,v_x), 
\\
\Phi_2(A_2) &=& (-A_2\,u_y-A_2^{\prime}\,x,-A_2\,v_y),
\\
\Phi_3(A_3) &=& (A_3, 0), 
\end{array}
\] 
while  the linear space  $\mathrm{cosym}_0(\EuScript{E}_2)$ of the second-order cosymmetries is spanned  by the functions 
\[
\begin{array}{lcl}
\psi_1 &=& (u, \Delta v),
\\
\psi_2 &=& (x^2+y^2,0),
\\
\Psi_1(B_1) &=& (B_1\,x,0),
\\
\Psi_2(B_2) &=& (B_2\,y,0),
\\
\Psi_3(B_3) &=& (B_3,0),
\\
\Upsilon_1(H_1) &=& (H_1,H_1^{\prime}\,\Delta u),
\\
\Upsilon_2(H_2) &=& (0, H_2),
\end{array}
\] 
where $A_i$ and $B_i$ are smooth functions of $t$  and $H_i$ are smooth functions of $v$. 
The action $\EuScript{H}\colon \mathrm{cosym}_0(\EuScript{E}_2) \rightarrow \mathrm{sym}_0(\EuScript{E}_2)$ 
is given by the formulas 
\[
\begin{array}{lcl}
\EuScript{H}(\psi_1) &=& -\varphi_4,
\\
\EuScript{H}(\psi_2) &=& 2\,\varphi_5,
\\
\EuScript{H}(\Psi_1(A_1)) &=& \Phi_1(A_1),
\\
\EuScript{H}(\Psi_2(A_2)) &=& -\Phi_2(A),
\\
\EuScript{H}(\Psi_3(A_3)) &=& \Phi_3(A_3^{\prime}),
\\
\EuScript{H}(\Upsilon_1(H_1)) &=& 0,
\\
\EuScript{H}(\Upsilon_2(H_2)) &=& 0.
\end{array}
\]
These formulas define the local expressions for the action of the inverse operator
$\EuScript{H}^{-1}$ on the symmetries $\varphi_4$, $\varphi_5$ and $\Phi_i(A_i)$, while the action
of $\EuScript{H}^{-1}$ on 
the symmetries $\varphi_k$ with $0 \le k \le 3$ can be described as follows.

Consider the four-parameter Lax representation 
\begin{equation}
\left\{
\begin{array}{lcl}
p_t &=& \J(u, p)+\mu_0\,\E(u),
\\
0 &=& \J(v, p)+\mu_0\,\E(v)+\mu_1\,v+\mu_2,
\\
q_t &=& \J(u, q) + \J(\Delta v, p)+\mu_0\,\E(\Delta v),
\\
0 &=& \J(v, q) + \J(\Delta u, p)+\mu_0\,\E(\Delta u)+\mu_1\,\Delta u+\mu_3
\end{array}
\right.
\label{2DiMHD_general_covering}
\end{equation}
of  \eqref{2DiMHD}. Then we have 
$\EuScript{H}^{-1}(\varphi_0) = (p_0, q_0)$, 
$\EuScript{H}^{-1}(\varphi_1) = (p_1-t\,u, q_1- t\,\Delta v)$,
$\EuScript{H}^{-1}(\varphi_2) = (p_2+\frac{1}{2}\,t\,(x^2+y^2), q_2)$,
and $\EuScript{H}^{-1}(\varphi_3) = (p_3, q_3)$, 
where $(p_0,q_0)$, $(p_1,q_1)$, $(p_2,q_2)$, and $(p_3,q_3)$ are solutions to system 
\eqref{2DiMHD_general_covering}
with    $(\mu_0, \mu_1,\mu_2,\mu_3)=(-1,0,0,0)$,
$(\mu_0, \mu_1,\mu_2,\mu_3)=(0,-1,0,0)$,
$(\mu_0, \mu_1,\mu_2,\mu_3)=(0,0,2,0)$,
and
$(\mu_0, \mu_1,\mu_2,\mu_3)=(0,0,0,1)$, respectively.

When $\mu_0 \neq 0$ in system \eqref{2DiMHD_general_covering}, we can put $\mu_0 = 1$ by the rescaling $(p,q) \mapsto (\mu_0\,p, \mu_0\,q)$. Similary, when $\mu_0=0$ and $\mu_1 \neq 0$ we can put $\mu_1=1$. 
We have the following sufficient conditions for the parameters $\mu_2$ and $\mu_3$ in 
\eqref{2DiMHD_general_covering} to be non-removable. 

\begin{theorem}
In system \eqref{2DiMHD_general_covering}
with $\mu_0 =0$ and $\mu_1 =1 $,  the parameter $\mu_2$ is non-removable, while the parameter $\mu_3$ is non-removable whenever $\mu_3 \neq 0$.  
\end{theorem}
\noindent
Proof.
Under the above assumptions,  the symmetries $\varphi_i$ with $i \in \{0,1,3\}$ do not admit lifts to symmetries of system
\eqref{2DiMHD_general_covering}.
Consider the vector fields $W_0 =x\,\partial_x+y\,\partial_y+2\,(u\,\partial_u+v\,\partial_v)$, 
$W_1 =-t\,\partial_t+u\,\partial_u+ v\,\partial_v$,
 and 
$W_3 = \partial_v$ which correspond to the generators $\varphi_0$, $-\varphi_1$,  and $\varphi_3$, respectively. 
Denote by $T_i(\varepsilon_i)$ the prolongations of the diffeomorphsims $\mathrm{exp}(\varepsilon\,W_i)$   to the bundle $J^3(\varpi)$ of the third-order jets of the sections of the bundle 
$\varpi \colon \mathbb{R}^7 \rightarrow \mathbb{R}^3$, $\varpi \colon (t,x,y,u,v,p,q) \mapsto (t,x,y)$.
Then the action of the diffeomorphism 
$T_3(\varepsilon_3) \circ T_0(\varepsilon_0)  \circ T_1(\varepsilon_1)$
and the rescaling $p \mapsto \mathrm{e}^{2\,\varepsilon_0}\,p$ map system \eqref{2DiMHD_general_covering}
to the same system with $\mu_2$ replaced by  
$\mathrm{e}^{2\,\varepsilon_0 +\varepsilon_1}\,(\mu_2 +\varepsilon_3)$
and $\mu_3$ replaced by $\mathrm{e}^{\varepsilon_1}\,\mu_3$.
Then the  assertion follows from \cite[\S\S~3.2, 3.6]{KrasilshchikVinogradov1989},
\cite{Krasilshchik2000,IgoninKrasilshchik2000,IgoninKerstenKrasilshchik2002,Marvan2002}.
\hfill $\Box$

\section{Concluding remarks}
In this paper we found new Lax representations \eqref{RMHD_first_covering}, \eqref{RMHD_second_covering}, \eqref{RMHD_third_covering}, \eqref{2DiMHD_covering}, \eqref{2DiMHD_second_covering}, and 
\eqref{2DiMHD_general_covering} for the reduced magnetohydrodynamics equations \eqref{RMHD} and the two-dimensional ideal 
magnetohydrodynamics equations \eqref{2DiMHD}. 
Systems \eqref{RMHD_first_covering} and \eqref{RMHD_second_covering}  provide new examples of Lax representations with a 
non-removable spectral parameter for a four-dimensional integrable system that is important from a physical point of view.
We construct B{\"a}cklund transformations between the tangent and the cotangent  coverings of the equations under the study.  
An important open question is to determine whether these B{\"a}cklund transformations define variational Poisson structures
for their equations.  
Another open problem is to ascertain whether the parameters in system \eqref{2DiMHD_general_covering} with $\mu_0=1$ are 
non-removable. 
Likewise, it is worth investigating whether other nonlocal structures --- such as nonlocal symmetries, nonlocal conservation laws,
and recursion operators --- are related to the obtained Lax representations.

\section*{Acknowledgments}

I would like to express my sincere gratitude  to I.S. Krasil${}^{\prime}$shchik  for insightful discussions.
I thank A.V. Samokhin for important remarks.

Computations  were done using the {\sc Jets} software \cite{Jets}.

\bibliographystyle{amsplain}

\end{document}